\documentclass[submission, Phys]{SciPost}
\usepackage[UKenglish]{babel}
\usepackage[utf8]{inputenc}
\usepackage[T1]{fontenc}

\usepackage{amssymb}
\usepackage{cancel}
\usepackage{nicefrac}
\usepackage{braket}
\usepackage[nameinlink,capitalize]{cleveref}
\usepackage{mathtools}

\newcommand{\tikzpic}[2]{
  \ifuseexttikz
  \includegraphics{#1}
  \else
  \tikzsetnextfilename{#1}
  #2
  \fi
}

\newcommand{\I}[0]{\mathrm{i}}

\newif\ifuseexttikz
\useexttikztrue

\ifuseexttikz
\else
\usepackage{pgfplots}
\usepackage{tikz}
\usetikzlibrary{external} 
\usetikzlibrary{shapes}
\usetikzlibrary{calc}
\usetikzlibrary{external}
\fi

\hypersetup{pdfauthor={Claudius Hubig, J. Ignacio Cirac},pdftitle={Time-dependent study of disordered models with infinite projected entangled pair states}}

\begin{document}

\begin{center}{\Large \textbf{
Time-dependent study of disordered models with infinite projected entangled pair states
}}\end{center}

\begin{center}
Claudius Hubig\textsuperscript{1,2,*} and J.~Ignacio Cirac\textsuperscript{1,2}
\end{center}

\begin{center}
{\bf 1} Max-Planck-Institut für Quantenoptik,\\
Hans-Kopfermann-Str. 1, 85748 Garching, Germany\\
{\bf 2} Munich Center for Quantum Science and Technology (MCQST),\\
Schellingstr. 4, 80799 München, Germany

{\bf *} claudius.hubig@mpq.mpg.de
\end{center}

\begin{center}
\today

\end{center}

\section*{Abstract} {\bf Infinite projected entangled pair states
  (iPEPS), the tensor network ansatz for two-dimensional systems in
  the thermodynamic limit, already provide excellent results on
  ground-state quantities using either imaginary-time evolution or
  variational optimisation. Here, we show (i) the feasibility of
  real-time evolution in iPEPS to simulate the dynamics of an infinite
  system after a global quench and (ii) the application of
  disorder-averaging to obtain translationally invariant systems in
  the presence of disorder. To illustrate the approach, we study the
  short-time dynamics of the square lattice Heisenberg model in the
  presence of a bi-valued disorder field. }

\section{Introduction}

Tensor networks are a powerful technique for the study of
strongly-correlated many-body problems. In particular, ground states
of local Hamiltonians are believed to be efficiently approximated by
tensor network states (TNS). In one dimension, the density matrix
renormalisation group (DMRG, \cite{white92:_densit}) and the
associated tensor network states called matrix-product states (MPS)
are the method of choice for the evaluation of ground-state
observables\cite{schollwoeck11}. In two dimensions, despite the worse
scaling of computational effort, infinite projected entangled pair
states (iPEPS, \cite{verstraete04:_renor,
  jordan08:_class_simul_infin_size_quant}), the two-dimensional
analogue of MPS, are already producing excellent
results, e.g.~Refs.~\cite{corboz10:_simul, corboz16:_improv_hubbar,
  corboz16:_variat}.

The study of dynamics using TNS is more difficult as the growth of the
entanglement entropy during a real-time evolution is typically
unfavourable. This usually limits the simulation of dynamics to
relatively short times. However, higher computational resources and
new time-evolution methods\cite{paeckel:_time} have made the study of
dynamics in one dimension commonplace with many interesting
applications, such as the calculation of excitation spectra, the use
as solvers in dynamical mean-field theory and the real-time simulation
of local and global quantum quenches.

In two dimensions, the situation is more difficult. Early studies of
time-evolution of a finite PEPS\cite{murg07:_variat} exist, but the
method has not seen wide-spread application. One reason is the lack of
gauge fixing in tensor networks with loops, which makes calculations
more costly and unstable. Additionally, most of the development was
focused on the infinite iPEPS ansatz which imposes translational
invariance and hence restricts the problems one may study.

In this paper we show how to study the dynamics of certain disordered
systems by simulating the time-evolution of an iPEPS. To make the
disordered problem translationally invariant, we employ the ancilla
trick introduced in
Ref.~\cite{paredes05:_exploit_quant_paral_simul_quant}. This trick
introduces ancilla spins on every site of the lattice. Coupling the
ancilla spins to the physical degrees of freedom and initialising the
ancillas in a superposition of spin-up and spin-down states leads to a
local superposition of discrete disorder configurations. The tensor
product of multiple sites results in an implicit averaging over all
possible discrete disorder realisations which is translationally
invariant and can be represented using iPEPS. To then simulate the
dynamics of this system, we use real-time evolution instead of
imaginary-time evolution while paying close attention to the numerical
stability of the ansatz and the errors incurred.

To illustrate the method, we simulate the dynamics of the XXX
Heisenberg model with a random bi-valued magnetic field; a similar
setup was already studied in one dimension using similar
techniques\cite{andraschko14:_purif_many_body_local_cold_atomic_gases}. While
we expect thermalisation to occur at sufficiently long times as
clusters with the same disorder field eventually couple, here we aim
at short time scales because, just as in one-dimensional MPS
simulations, we are limited in the times we can obtain due to the
entanglement growth in the system. However, even short-time
simulations are of interest when comparing to recent
experiments\cite{choi16:_explor} which likewise study the real-time
dynamics of two-dimensional disordered systems. Our intent is then to
find qualitative effects of disorder on such short time scales.

To this aim, the rest of the paper is structured as follows:
\cref{sec:ancilla} revisits the ancilla trick to incorporate the
disorder averaging. \cref{sec:hinf} introduces our physical test model
and discusses the expected behaviour at infinite disorder
strength. \cref{sec:timeevo} describes the steps taken to obtain
reliable real-time data within the iPEPS framework. Finally,
\cref{sec:results} discusses our results obtained from the disordered
two-dimensional Heisenberg model before drawing conclusions in
\cref{sec:conclusions}.

\section{\label{sec:ancilla}Averaging via auxiliary spins}

We use infinite projected entangled pair
states\cite{jordan08:_class_simul_infin_size_quant} to study the
two-dimensional system in the thermodynamic limit. The tensor network
ansatz is based on a finite unit cell of tensors repeated infinitely
whose (ideal) contraction represents the infinite quantum state of
(e.g.) physical $S=\nicefrac{1}{2}$ spins. This ansatz requires some
kind of translational invariance of the state e.g. under translation
of 2 sites to the left. Such translational invariance is incompatible
with randomly disordered systems. To avoid this problem, we restrict
the disorder to be discrete and bi-valued, that is, for any disorder
configuration $\mathcal{A}$, the function $z_{i;\mathcal{A}}$ assigns
each lattice site a value $\pm \nicefrac{1}{2}$. We then follow
Ref.~\cite{paredes05:_exploit_quant_paral_simul_quant} and introduce
ancilla $S=\nicefrac{1}{2}$ spins on each lattice site $i$, doubling
the local space (with $p$ and $a$ denoting physical and ancilla space
respectively):
\begin{equation}
  \ket{\sigma_i} \to \ket{\sigma_i}_p \otimes \ket{\sigma_i}_a \;.
\end{equation}
We then initialise all ancilla spins in the $\ket{+}$ state
\begin{equation}
  \ket{+_i}_a = \frac{1}{\sqrt{2}}\left(\ket{\uparrow_i}_a + \ket{\downarrow_i}_a\right) \; \forall i
\end{equation}
resulting in an initial ancilla state of $N$ sites
\begin{align}
  \ket{+}_a & = \left(\ket{+_i}_a\right)^{\otimes N} \\
            & = \frac{1}{\sqrt{2^N}}\sum_{\sigma_i = \uparrow,\downarrow} \ket{\sigma_1 \sigma_2 \ldots \sigma_N}_a \\
            & \equiv \frac{1}{\sqrt{2^N}} \sum_{\substack{\textrm{\tiny{disorder}}\\\textrm{\tiny{config.}}\;\tiny{\mathcal{A}}}} \ket{\mathcal{A}}_a \;.
\end{align}
For example, the tensor product of e.g.~$3$ ancilla spins is the state
\begin{align}
  \ket{+_1}_a \otimes \ket{+_2}_a \otimes \ket{+_3}_a = & \frac{1}{\sqrt{8}} \left(\ket{\uparrow_1}_a + \ket{\downarrow_1}_a\right) \otimes \left(\ket{\uparrow_2}_a + \ket{\downarrow_2}_a\right) \otimes \left(\ket{\uparrow_3}_a + \ket{\downarrow_3}_a\right) \\
                                                         = & \frac{1}{\sqrt{8}} \Big( \ket{\uparrow_1 \uparrow_2 \uparrow_3}_a + \ket{\uparrow_1 \uparrow_2 \downarrow_3}_a + \ket{\uparrow_1 \downarrow_2 \uparrow_3}_a + \ket{\uparrow_1 \downarrow_2 \downarrow_3}_a \\
                                                         & \qquad + \ket{\downarrow_1 \uparrow_2 \uparrow_3}_a + \ket{\downarrow_1 \uparrow_2 \downarrow_3}_a + \ket{\downarrow_1 \downarrow_2 \uparrow_3}_a + \ket{\downarrow_1 \downarrow_2 \downarrow_3}_a \Big) \;,
\end{align}
i.e.~an equal superposition of all possible $\uparrow,\downarrow$
configurations.
We now consider the physical Hamiltonian for a specific disorder
realisation $\mathcal{A}$
\begin{equation}
  \hat H_{\mathcal{A};p} = \sum_{\braket{i,j}} \hat s_{i;p} \cdot \hat s_{j;p} + h \sum_i \hat s^z_{i;p} z_{i;\mathcal{A}}
\end{equation}
where $\hat s_{i;p}$ are physical spin operators, $\hat s^z$ denotes
the $z$-component of the spin operator only, $\braket{i,j}$ denotes
nearest-neighbour sites $i=x,y$ and $j=x,y\pm 1$ or $j=x\pm 1,y$, the
effective disorder field has a strength $h \geq 0$ and the function
$z_{i;\mathcal{A}}$ assigns each lattice site a value
$\pm \nicefrac{1}{2}$. Now the time-dependent disorder-averaged
expectation value of $\hat O_p$ starting from some initial physical
state $\ket{\psi}$ is
\begin{align}
  \langle\!\langle \hat O_p(t) \rangle\!\rangle & \equiv \frac{1}{2^N} \sum_{\substack{\textrm{\tiny{disorder}}\\\textrm{\tiny{config.}}\;\tiny{\mathcal{A}}}} \prescript{}{p}{\bra{\psi}} \hat O(t) \ket{\psi}_p \\
                                                & = \frac{1}{2^N} \sum_{\substack{\textrm{\tiny{disorder}}\\\textrm{\tiny{config.}}\;\tiny{\mathcal{A}}}} \prescript{}{p}{\bra{\psi}} e^{\I t \hat H_{\mathcal{A};p}} \hat O e^{-\I t \hat H_{\mathcal{A};p}}  \ket{\psi}_p \\
                                                & = \frac{1}{2^N} \sum_{\substack{\textrm{\tiny{disorder}}\\\textrm{\tiny{config.}}\;\tiny{\mathcal{A}}}} \left( \prescript{}{a}{\bra{\mathcal{A}}} \otimes \prescript{}{p}{\bra{\psi}} \right) e^{\I t \hat H_{\mathcal{A};p}} \hat O e^{-\I t \hat H_{\mathcal{A};p}}  \left( \ket{\psi}_p \otimes \ket{\mathcal{A}}_a \right) \\
                                                & = \frac{1}{2^N} \sum_{\substack{\textrm{\tiny{disorder}}\\\textrm{\tiny{config.}}\;\tiny{\mathcal{A}}}} \left( \prescript{}{a}{\bra{\mathcal{A}}} \otimes \prescript{}{p}{\bra{\psi}} \right) e^{\I t \hat H} \hat O e^{-\I t \hat H}  \left( \ket{\psi}_p \otimes \ket{\mathcal{A}}_a \right) \\
                                                & = \left( \prescript{}{a}{\bra{+}} \otimes \prescript{}{p}{\bra{\psi}} \right) e^{\I t \hat H} \hat O e^{-\I t \hat H}  \left( \ket{\psi}_p \otimes \ket{+}_a \right)
\end{align}
with the effective Hamiltonian
\begin{equation}
  \hat H = \sum_{\braket{i,j}} \hat s_{i;p} \cdot \hat s_{j;p} + h \sum_i \hat s^z_{i;p} \hat s^z_{i;a} \label{eq:hamiltonian}
\end{equation}
and $\hat s_{i;a}$ the auxiliary spin operator on site $i$.

\section{\label{sec:hinf}The Heisenberg model with discrete disorder}

The Hamiltonian \cref{eq:hamiltonian} and the initial state
$\ket{\psi}_a$ of the ancilla spins is translationally invariant. We
initialise the physical spins in a Néel state (i.e., checkerboard
pattern), which can be described by a $2 \times 2$ unit cell.

In this situation, at $h = \infty$, dynamics are constrained to
clusters of the same disorder configuration. Those clusters are
randomly sized and shaped and will support no dynamics ($1 \times 1$
``cluster''), some oscillations (small clusters) or essentially
thermodynamic-limit like behaviour (for very large clusters). In any
specific disorder realisation, any particular site has probability
$p = \nicefrac{1}{2}$ of belonging to the ancilla-up or ancilla-down
type. This probability is below the critical percolation probability
$p_c \approx 0.592$\cite{berg96, gebele84:_site} of the square lattice
site percolation problem\cite{grimmett99:_percol}. As a result, there
is no cluster of infinite size and the probability of a given cluster
occurring decays exponentially with its size. Because odd cluster sizes
are more common than even cluster sizes (1 is more common than 2, 3 is
more common than 4 etc.), the sign of the average magnetisation of a
given cluster even at long times will coincide with the initial
magnetisation of the chosen site.

A first extension of the present study would be $S=1$ auxiliary
ancilla spins. There, too, cluster probability decays exponentially in
cluster size, but this decay is much faster. One hence expects
stronger localisation effects and less pronounced oscillatory
dynamics. In the limit of infinite local ancilla spins, each cluster
will have size $1$ and the system will hence be completely frozen
(still at $h = \infty$), corresponding to a standard model with
infinitely strong continuous disorder.

Going away from $h = \infty$ to smaller disorder strengths then leads
to some coupling between clusters. Since clusters with equal
configurations of ancilla spins can couple through potential barriers
of opposite-orientation ancilla spins at times linear in the disorder
strength, we expect eventual equilibration for discrete disorder
configurations as generated by finite ancilla spins. However, even if
equilibration occurs eventually, we expect to see an effect in the
short-term dynamics after a global quench.

\section{\label{sec:timeevo}Real-time evolution in iPEPS}

Real-time evolution of an iPEPS is ``technically
straightforward''\cite{murg07:_variat, phien15:_fast}, it suffices to
replace the standard prefactor $\beta$ used during imaginary-time
evolution to find the ground state by a real time step $\I t$. In
practice, the quick growth of entanglement and instability of the
iPEPS is problematic. Here, we are using a second-order Trotter-Suzuki
decomposition\cite{suzuki76:_gener_trott} of the Hamiltonian
\cref{eq:hamiltonian}. There are four bonds inside the $2 \times 2$
unit cell and four bonds linking each unit cell to its right and upper
neighbour. With the symmetrisation due to the second-order
decomposition, we hence have to apply 16 individual gates to perform a
single time step. The innermost two gates can be combined to reduce
the gate count by one. After each individual gate application, we then
use the full update\cite{jordan08:_class_simul_infin_size_quant} to
recompute the environment for maximal accuracy. During the simulation,
we use the $\mathrm{U}(1)$-$S^z$ symmetry of the system in the
physical sector\cite{hubig18:_abelian} which allows us to take the
bond dimension up to $D = 9$. Each site of the iPEPS has
local dimension $4$ as we combine physical and ancilla spins into
individual sites. The initial N\'eel state is represented by an iPEPS
with bond dimension $D=1$.

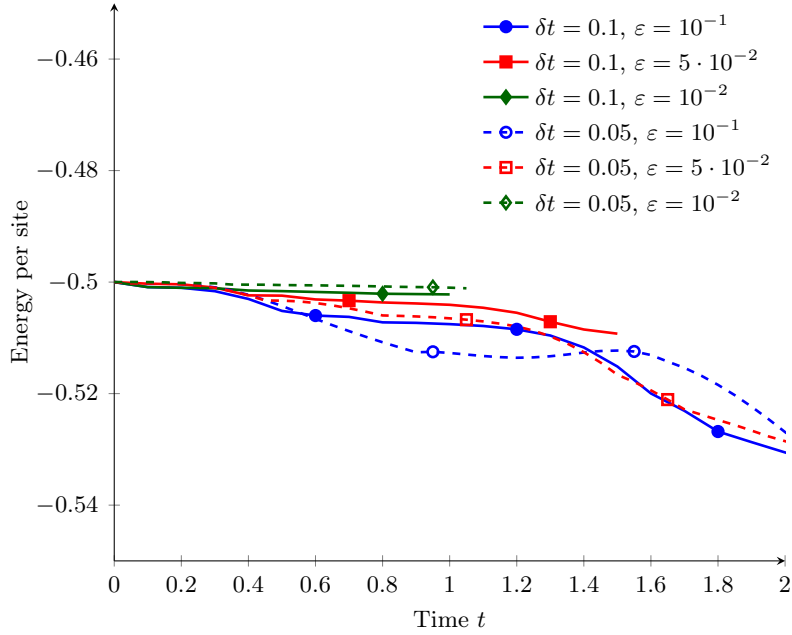
\begin{figure}[p]
  \centering
  \tikzpic{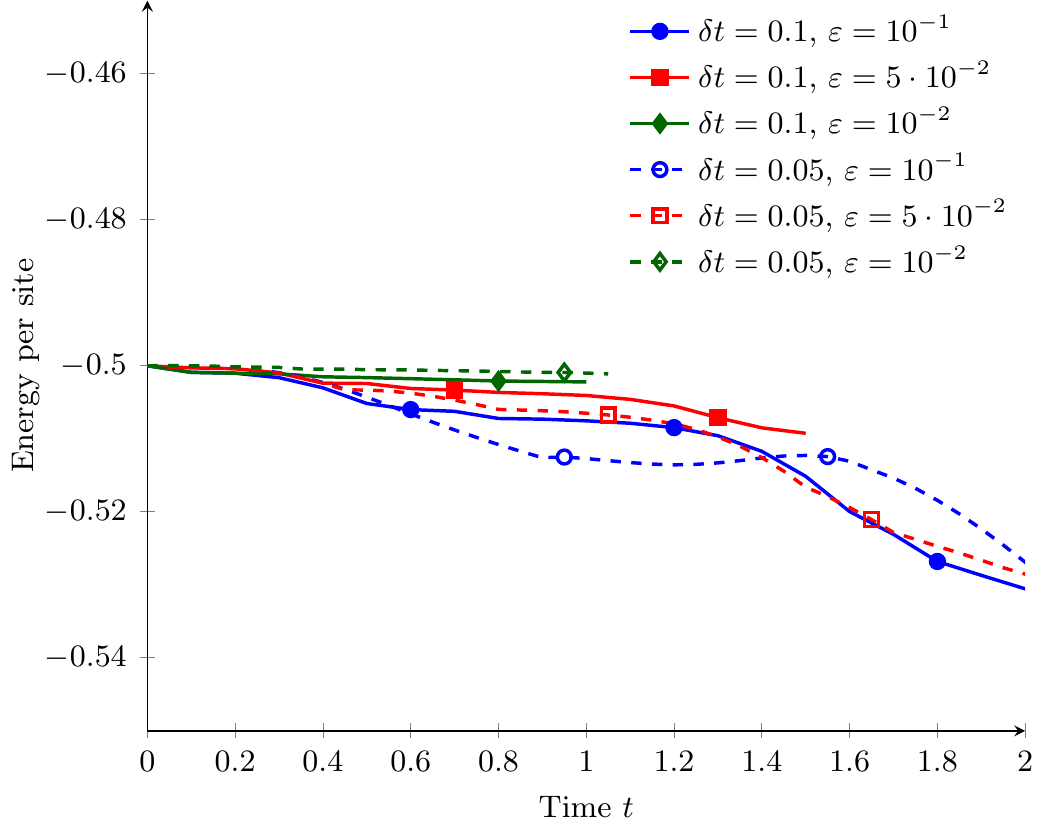}{
    \pgfplotsset{every axis plot/.append style={
        line width=1pt,
        tick style={line width=0.6pt}}}
    \begin{tikzpicture}
      \begin{axis}[
        width=0.7\textwidth,
        height=0.6\textwidth,
        axis lines=left,
        xlabel={Time $t$},
        font = {\footnotesize},
        ylabel={Energy per site},
        xmin=0,
        xmax=2,
        ymin=-0.55,
        ymax=-0.45,
        legend style = {draw=none,fill=none,anchor=north east, at = {(1,1)}},
        legend cell align = left,
        legend columns=1,
        legend image post style = {scale=1},
        ]
        \addplot[color=blue,solid,mark=*,mark repeat=6,mark phase=7]table[y expr = \thisrowno{1}/4]{data/4/4-1e-1-0.1_energy};
        \addlegendentry{$\delta t = 0.1$, $\varepsilon = 10^{-1}$};
        \addplot[color=red,solid,mark=square*,mark repeat=6,mark phase=8]table[y expr = \thisrowno{1}/4]{data/4/4-5e-2-0.1_energy};
        \addlegendentry{$\delta t = 0.1$, $\varepsilon = 5\cdot 10^{-2}$};
        \addplot[color=green!40!black,solid,mark=diamond*,mark repeat=6,mark phase=9,mark options={scale=1.2}]table[y expr = \thisrowno{1}/4]{data/4/4-1e-2-0.1_energy};
        \addlegendentry{$\delta t = 0.1$, $\varepsilon = 10^{-2}$};
        \addplot[color=blue,dashed,mark=o,mark repeat=12,mark options=solid,mark phase=20]table[y expr = \thisrowno{1}/4]{data/4/4-1e-1-0.05_energy};
        \addlegendentry{$\delta t = 0.05$, $\varepsilon = 10^{-1}$};
        \addplot[color=red,dashed,mark=square,mark repeat=12,mark phase=22,mark options=solid]table[y expr = \thisrowno{1}/4]{data/4/4-5e-2-0.05_energy};
        \addlegendentry{$\delta t = 0.05$, $\varepsilon = 5\cdot 10^{-2}$};
        \addplot[color=green!40!black,dashed,mark=diamond,mark repeat=12,mark phase=20,mark options={solid, scale=1.2}]table[y expr = \thisrowno{1}/4]{data/4/4-1e-2-0.05_energy};
        \addlegendentry{$\delta t = 0.05$, $\varepsilon = 10^{-2}$};
      \end{axis}
    \end{tikzpicture}
  }
  \caption{\label{fig:h4:e}Energy per site over time at $h=4$ as
    obtained from some stable example configurations for $\delta t$
    and $\varepsilon$. For the later plots \cref{fig:e,fig:sz}, we
    have selected $\delta t = 0.1$, $\varepsilon = 0.05$ (solid red
    squares) to represent $h = 4$ as the error is sufficiently
    small.}
\end{figure}
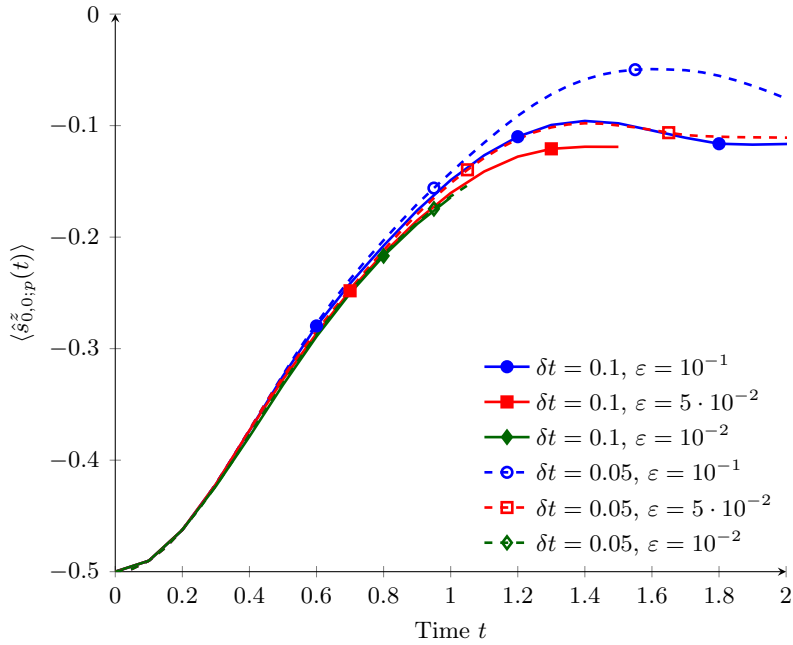
\begin{figure}[p]
  \centering
    \tikzpic{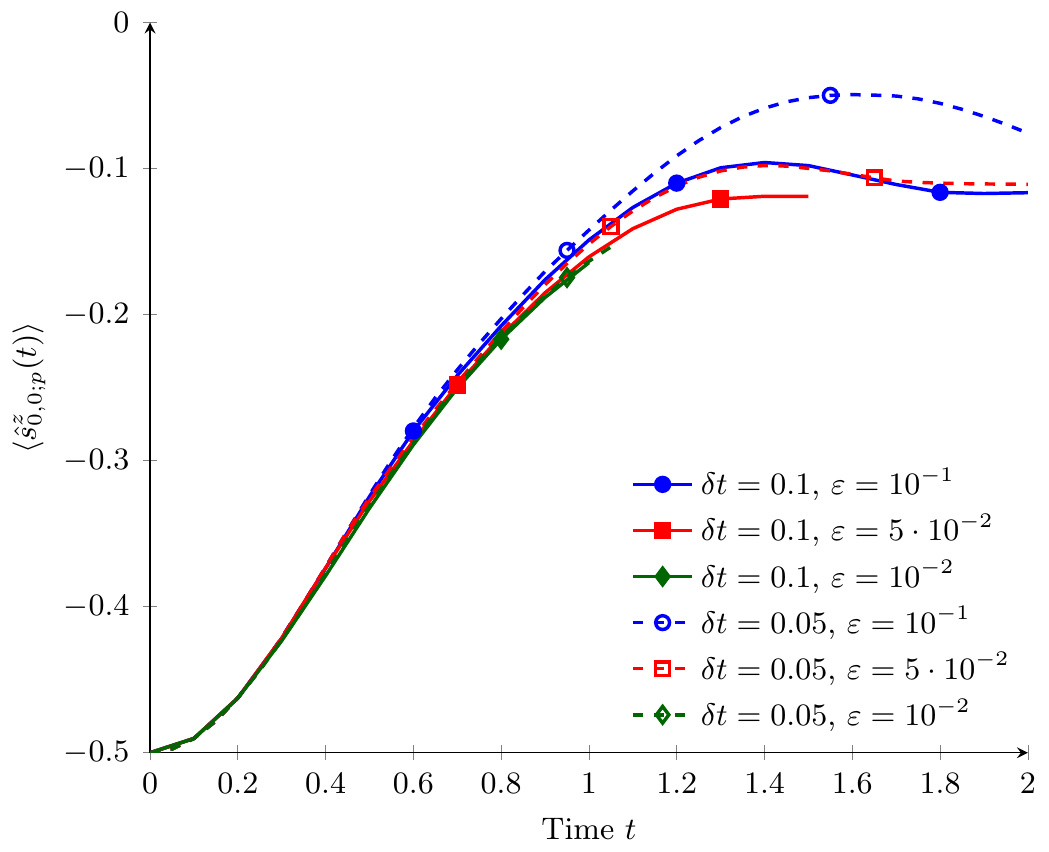}{
      \pgfplotsset{every axis plot/.append style={
          line width=1pt,
          tick style={line width=0.6pt}}}
      \begin{tikzpicture}
        \begin{axis}[
          width=0.7\textwidth,
          height=0.6\textwidth,
          axis lines=left,
          xlabel={Time $t$},
          font = {\footnotesize},
          ylabel={$\braket{\hat s^z_{0,0;p}(t)}$},
          xmin=0,
          xmax=2,
          ymin=-0.5,
          ymax=0,
          legend style = {draw=none,fill=none,anchor=south east, at = {(1,0.01)}},
          legend cell align = left,
          legend columns=1,
          legend image post style = {scale=1},
          ]
          \addplot[color=blue,solid,mark=*,mark repeat=6,mark phase=7]table[y expr = \thisrowno{1}]{data/4/4-1e-1-0.1_sz};
          \addlegendentry{$\delta t = 0.1$, $\varepsilon = 10^{-1}$};
          \addplot[color=red,solid,mark=square*,mark repeat=6,mark phase=8]table[y expr = \thisrowno{1}]{data/4/4-5e-2-0.1_sz};
          \addlegendentry{$\delta t = 0.1$, $\varepsilon = 5\cdot 10^{-2}$};
          \addplot[color=green!40!black,solid,mark=diamond*,mark repeat=6,mark phase=9,mark options={scale=1.2}]table[y expr = \thisrowno{1}]{data/4/4-1e-2-0.1_sz};
          \addlegendentry{$\delta t = 0.1$, $\varepsilon = 10^{-2}$};
          \addplot[color=blue,dashed,mark=o,mark repeat=12,mark options=solid,mark phase=20]table[y expr = \thisrowno{1}]{data/4/4-1e-1-0.05_sz};
          \addlegendentry{$\delta t = 0.05$, $\varepsilon = 10^{-1}$};
          \addplot[color=red,dashed,mark=square,mark repeat=12,mark phase=22,mark options=solid]table[y expr = \thisrowno{1}]{data/4/4-5e-2-0.05_sz};
          \addlegendentry{$\delta t = 0.05$, $\varepsilon = 5\cdot 10^{-2}$};
          \addplot[color=green!40!black,dashed,mark=diamond,mark repeat=12,mark phase=20,mark options={solid, scale=1.2}]table[y expr = \thisrowno{1}]{data/4/4-1e-2-0.05_sz};
          \addlegendentry{$\delta t = 0.05$, $\varepsilon = 10^{-2}$};
        \end{axis}
      \end{tikzpicture}
    }
    \caption{\label{fig:h4:sz}Expectation value of
      $\braket{\hat s^z_{0,0;p}(t)}$ at $h=4$ for the calculations
      also shown in \cref{fig:h4:e}. The selected setting
      $\delta t = 0.1$, $\varepsilon = 0.05$ is very close to the
      higher-precision calculations with $\varepsilon = 10^{-2}$,
      though those exhaust the available bond dimension earlier.}
\end{figure}

Contrary to MPS, iPEPS cannot be gauged\cite{lubasch14:_algor,
  phien15:_fast, phien15:_infin} perfectly. Using a bond dimension $D$
in an iPEPS to represent a state which would only require a much
smaller bond dimension $D^\prime \ll D$ leads to serious problems with
numerical stability in both the corner-matrix renormalisation group
procedure as well as the next full update due to a badly-conditioned
norm tensor. As entanglement grows, we need to increase the bond
dimension during some time steps from $D$ to $D+1$ and hence have to
balance multiple objectives: First, small time steps lead to a small
Trotter error, which is desired. Second, large time steps allow
increasing the bond dimension in a very stable manner, as the change
induced by the step is large and hence additional entanglement between
the two involved sites is created. Third, even in a series of large
time steps, we cannot grow the bond dimension at every step and hence
have to select the optimal step at which numerical stability is
ensured and the loss of information due to too strong a truncation is
minimised. Hence increasing the bond dimension too quickly leads to
problems with numerical stability, increasing it too late constrains
the dynamics to a lowly-entangled subspace. In an imaginary-time
evolution to find the ground state, one can typically ignore this
problem and accept some numerical instability for a few steps after
increasing $D$. Here, we do not have that luxury and need to be
precise and stable during all steps.

Overall, there are three free parameters in our calculation: first,
the step size $\delta t$ used in the Trotter decomposition. We select
$\delta t = 0.1, 0.05, 0.025$ and $0.01$ and check that we obtain
relatively converged results between those. Second, we measure the
accumulated cost of truncation during a full time step as the sum of
the norms of differences between truncated and untruncated states at
each individual gate application and, once this cost crosses a
threshold $\varepsilon$, increase the bond dimension $D$ by 1. We
select $\varepsilon = 10^{-1,-2,-3,-4}$ and $5\cdot10^{-2}$. Finally,
the physical parameter $h$ couples the ancilla spins to the physical
spins. We select $h = 0, 1, 4, 16, 64, 256$. The environment bond
dimension $\chi$ is set to $10(D+1)$ throughout the evolution, testing
suggests that this is sufficient.

\subsection{Converging results in $\delta t$ and $\varepsilon$}

For fixed $h$, we then compare the results at different $\varepsilon$
and $\delta t$. The evaluated expectation values are $\braket{\hat H}$
and $\braket{\hat s^z_{0,0;p}}$ on the first unit cell site. The spins
on the other three unit cell sites behave as expected as
$\pm\braket{\hat s^z_{0,0;p}}$ to very good accuracy.

Some of the calculations are still unstable
(e.g.~$\varepsilon = 10^{-4}$ tends to be problematic) leading to a
very abrupt and large error in the energy; this problem is more
pronounced at larger values of $h$ and smaller step sizes
$\delta$. The instability is due to an ``insufficient'' growth of
entanglement in the iPEPS which does not fill up the allowed bond
dimension $D$. This situation can be diagnosed reliably and the
associated calculations are discarded.

Furthermore, if $\varepsilon$ is too large and the time step
$\delta t$ is too small, a large error is made at every step which
over time leads to no entanglement growth (even at $h=0$!) and instead
to wrong oscillatory dynamics over a limited range. Diagnosing this
problem based only on the energy and spin expectation values is very
difficult, as the energy is conserved relatively very well and the
spin expectation values form a smooth curve. We instead focus on the
growth of the bond dimension $D$. If during the calculation the bond
dimension does not grow to the maximal value $D = 9$, we assume that the
truncation error made is too large and the calculation is discarded.

The above two criteria discard a number of calculations, especially at
small step sizes and very large or very small error threshold. From
the rest, we obtain fairly well-converged data typically at
$\delta t \leq 0.05$ and $\varepsilon = 10^{-2}$. Figs.~\ref{fig:h4:e}
and \ref{fig:h4:sz} show exemplary data curves at $h = 4$ once the
unstable and wrong results were removed. We obtain good convergence of
the local observable $\braket{\hat s^z_{0,0;p}(t)}$ for a range of
time steps $\delta t = 0.1, 0.05$ and error thresholds
$\varepsilon = 10^{-1}, 5\cdot 10^{-2}, 10^{-2}$ and $10^{-3}$ (not
shown). One calculation at $\varepsilon = 10^{-1}$ and $\delta = 0.05$
produces very different dynamics from the others likely due to the
large truncation error. Decreasing the truncation error at fixed time
step size or increasing the time step size at fixed truncation error
per step leads to a number of very similar results.

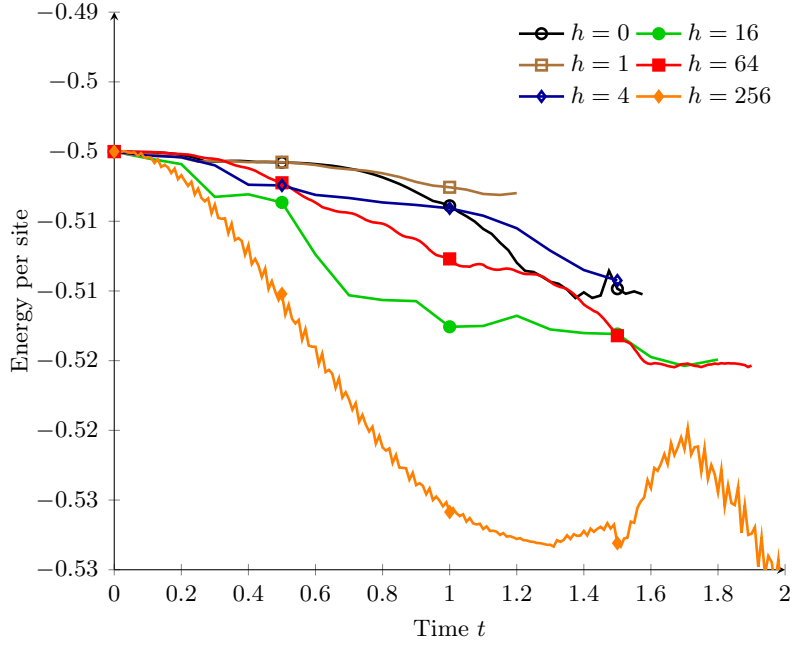
\begin{figure}[p]
  \centering
  \tikzpic{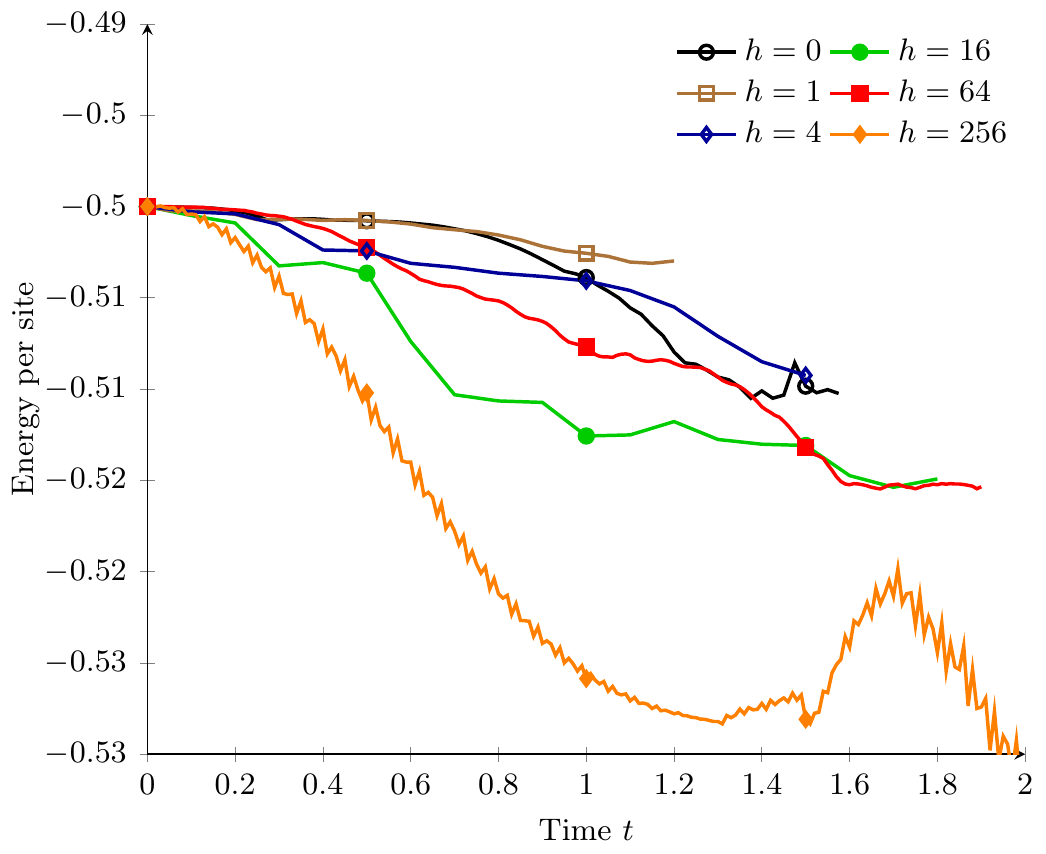}{
    \pgfplotsset{every axis plot/.append style={
        line width=1pt,
        tick style={line width=0.6pt}}}
    \begin{tikzpicture}
      \begin{axis}[
        width=0.7\textwidth,
        height=0.6\textwidth,
        axis lines=left,
        xlabel={Time $t$},
        font = {\footnotesize},
        ylabel={Energy per site},
        xmin=0,
        xmax=2,
        ymin=-0.53,
        ymax=-0.49,
        legend style = {draw=none,fill=none,anchor=north east, at = {(1,1)}},
        legend cell align = left,
        legend columns=2,
        legend image post style = {scale=1},
        ]
        \addplot[color=black,solid,mark=o,mark repeat=20]table[y expr = \thisrowno{1}/4]{data/0/0-1e-2-0.025_energy};
        \addlegendentry{$h = 0$};

        \addplot[color=green!80!black,solid,mark=*,mark repeat=5]table[y expr = \thisrowno{1}/4]{data/16/16-5e-2-0.1_energy};
        \addlegendentry{$h = 16$};

        \addplot[color=brown!90!black,solid,mark=square,mark repeat=10,mark options=solid]table[y expr = \thisrowno{1}/4]{data/1/1-1e-2-0.05_energy};
        \addlegendentry{$h = 1$};

        \addplot[color=red,solid,mark=square*,mark repeat=50,mark options=solid]table[y expr = \thisrowno{1}/4]{data/64/64-1e-2-0.01_energy};
        \addlegendentry{$h = 64$};

        \addplot[color=blue!60!black,solid,mark=diamond,mark repeat=5,mark options=solid]table[y expr = \thisrowno{1}/4]{data/4/4-5e-2-0.1_energy};
        \addlegendentry{$h = 4$};

        \addplot[color=orange,solid,mark=diamond*,mark options=solid,mark repeat=50]table[y expr = \thisrowno{1}/4]{data/256/256-1e-2-0.01_energy};
        \addlegendentry{$h = 256$};
      \end{axis}
    \end{tikzpicture}
  }
  \caption{\label{fig:e}Energy per site over time at various disorder
    strengths. In all calculations, the error in energy is relatively
    small and typically smaller than the accumulated cost of
    truncation.}
\end{figure}

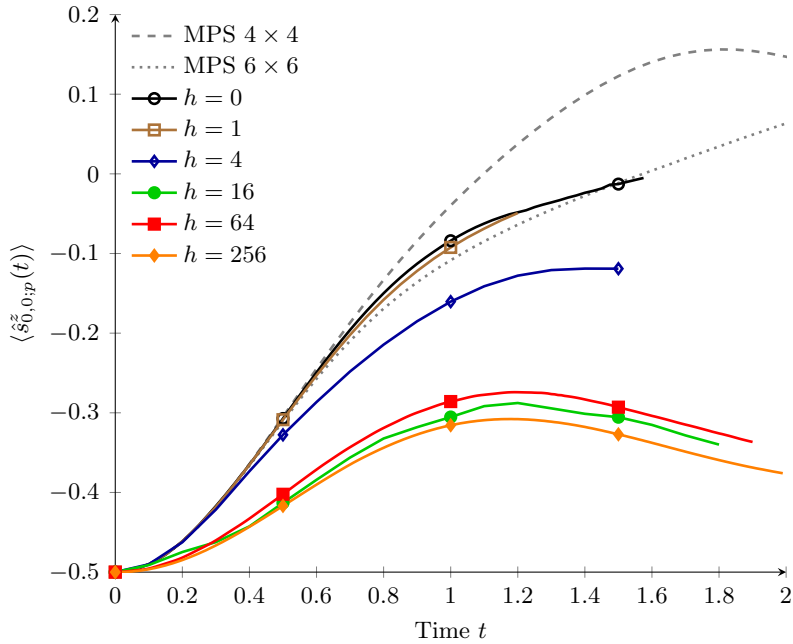
\begin{figure}[p]
  \centering
  \tikzpic{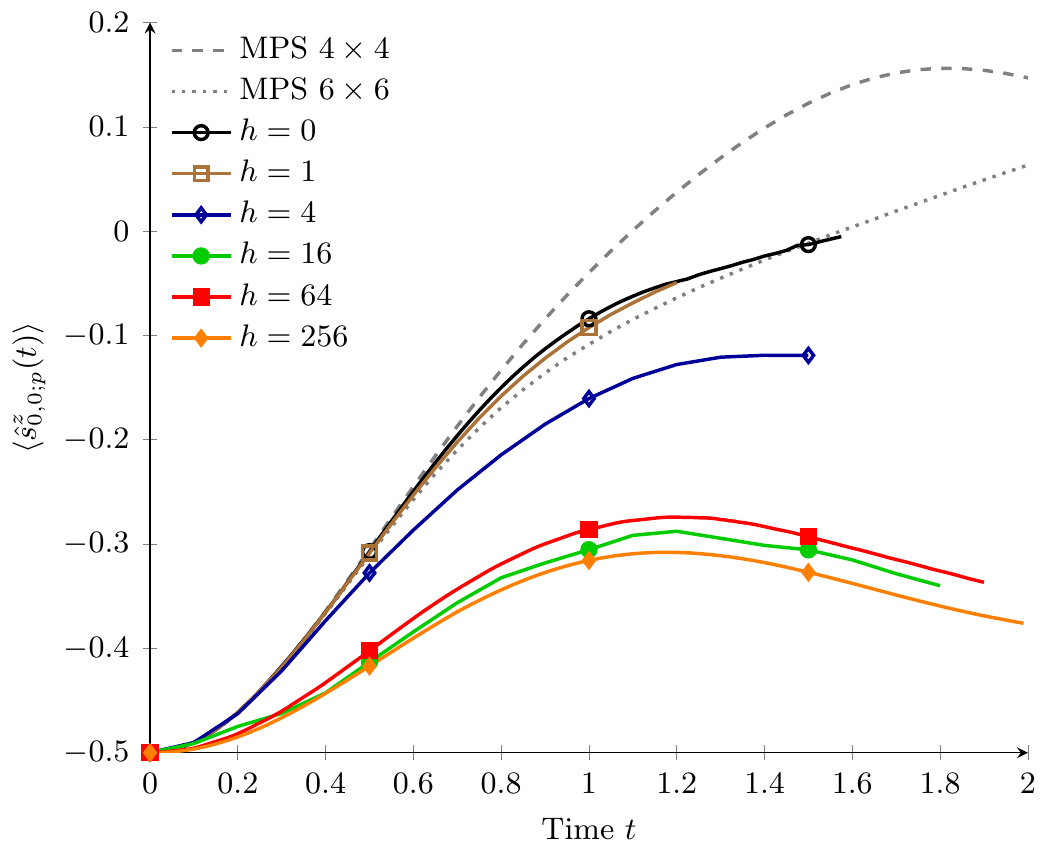}{
    \pgfplotsset{every axis plot/.append style={
        line width=1pt,
        tick style={line width=0.6pt}}}
    \begin{tikzpicture}
      \begin{axis}[
        width=0.7\textwidth,
        height=0.6\textwidth,
        axis lines=left,
        xlabel={Time $t$},
        font = {\footnotesize},
        ylabel={$\braket{\hat s^z_{0,0;p}(t)}$},
        xmin=0,
        xmax=2,
        ymin=-0.5,
        ymax=0.2,
        legend style = {draw=none,fill=none,anchor=north west, at = {(0.01,1)}},
        legend cell align = left,
        legend columns=1,
        legend image post style = {scale=1},
        ]
        \addplot[color=gray,dashed]table[y expr = \thisrowno{1}]{data/mps-4x4};
        \addlegendentry{MPS $4 \times 4$};
        \addplot[color=gray,dotted]table[y expr = \thisrowno{1}]{data/mps-6x6};
        \addlegendentry{MPS $6 \times 6$};
        \addplot[color=black,solid,mark=o,mark repeat=20]table[y expr = \thisrowno{1}]{data/0/0-1e-2-0.025_sz};
        \addlegendentry{$h = 0$};

        \addplot[color=brown!90!black,solid,mark=square,mark repeat=10,mark options=solid]table[y expr = \thisrowno{1}]{data/1/1-1e-2-0.05_sz};
        \addlegendentry{$h = 1$};

        \addplot[color=blue!60!black,solid,mark=diamond,mark repeat=5,mark options=solid]table[y expr = \thisrowno{1}]{data/4/4-5e-2-0.1_sz};
        \addlegendentry{$h = 4$};

        \addplot[color=green!80!black,solid,mark=*,mark repeat=5]table[y expr = \thisrowno{1}]{data/16/16-5e-2-0.1_sz};
        \addlegendentry{$h = 16$};

        \addplot[color=red,solid,mark=square*,mark repeat=50,mark options=solid]table[y expr = \thisrowno{1}]{data/64/64-1e-2-0.01_sz};
        \addlegendentry{$h = 64$};

        \addplot[color=orange,solid,mark=diamond*,mark options=solid,mark repeat=50]table[y expr = \thisrowno{1}]{data/256/256-1e-2-0.01_sz};
        \addlegendentry{$h = 256$};
      \end{axis}
    \end{tikzpicture}
  }
  \caption{\label{fig:sz}Expectation value of
    $\braket{\hat s^z_{0,0;p}(t)}$ at various disorder strengths. A
    noticeable slow-down of dynamics is observed at increased disorder
    strengths $h \geq 16$; MPS calculations at $h = 0$ show
    finite-size effects at $t < 1$.}
\end{figure}

\section{\label{sec:results}Disorder-averaged dynamics}

Using the procedure above, we obtain one converged curve
$\braket{\hat s^z_{0,0;p}(t)}$ per disorder coupling strength
$h$. Depending on the entanglement growth in the system and the
threshold $\varepsilon$ selected, the calculations exceed the
available resources at different times $t$. As a result, not all
calculations reach our maximal target time $t = 2$. \cref{fig:e} shows
the energy per site in these calculations. We observe the energy to be
conserved relatively well with an error comparable in magnitude to the
cost function measured during each iPEPS full update, the relative
error in energy per site is never higher than a few per cent even
after many full-update steps. We do not take any particular steps to
enforce energy conservation during the update. The error in energy can
hence be -- to some extend -- considered an error measure for our
calculation itself. As a side effect of the Trotterisation, in
particular the energy conservation is affected by the Trotter
error. It is then understandable that at larger disorder strengths
$h$, the commutator between the local field terms
$h \, \hat s^z_{i;p} \hat s^z_{i;a}$ and bond terms
$\hat s^x_{i;p} \hat s^x_{j;p}$ necessarily becomes larger and we
observe worse energy conservation.

In \cref{fig:sz} we analyse the time-dependent expectation value
$\braket{\hat s^z_{0,0;p}(t)}$. Very weak disorder $h = 1$ leads to
essentially the same dynamics as the case without disorder ($h = 0$).
At $h = 4$, the initial dynamics ($t \leq 0.5$) are still comparable
to those of the clean system and only later times ($t \approx 1$) show
an effect due to the disorder. Increasing the disorder strength
further we find distinctly different dynamics at $h \geq 16$. These
differences set on very early ($t \approx 0.1$) and lead to noticeably
weaker oscillations, with the first peak at $t \approx 1$ at
$\braket{\hat s^z_{0,0;p}(1)} \approx -0.3$. Of course, we would
expect that stronger disorder leads to monotonically increasing
differences in the dynamics, which does not appear to be the case in
\cref{fig:sz}. The obtained data range at $h = [16, 64, 256]$ may
hence also serve as an approximate error measure for our
calculation. Nevertheless, a drastic difference to data at $h \leq 4$
is visible, both in the very precise shortest-time frame $t \leq 0.5$
and at longer times $t \geq 1$.

By comparing with time-dependent MPS calculations\cite{paeckel19:_time}
on $4 \times 4$ and $6 \times 6$ tori in the zero-disorder case, we
find finite-size effects at time $t > 0.5$. Comparable or smaller
system sizes are available to exact diagonalisation studies while numerical linked cluster calculations on similar systems\cite{white17:_quant} likewise encounter finite-size effects around $t < 1$. Our
method, extending the time frame to at least $t \approx 1.5$, is hence
certainly an improvement over alternative MPS-based
approaches. Additionally, future improvements to increase the
stability of the calculation and further increase the bond dimension
should help in obtaining longer times without having to worry about
finite-size effects.

\section{\label{sec:conclusions}Conclusion}

Presently, the iPEPS formalism allows for the simulation of the
real-time evolution of quantum states free from boundary
effects. While limited in the obtainable times due to entanglement
growth in the present system, this limit is still beyond the onset of
finite-size effects in comparable matrix-product state, exact
diagonalisation or numerical linked cluster approximation
calculations. Using the formalism, we were able to study a
discrete-disorder-averaged two-dimensional Heisenberg model, where we
have found a slowdown of dynamics at disorder coupling strengths
$h \geq 16$. This slowdown is markedly different from the dynamics
encountered in the zero-disorder case. The generalisation to
multi-valued disorder configurations will be the topic of future work.

The two primary challenges at the moment are the stability of the full
update, which limits the obtainable precision by requiring a certain
minimal cost of truncation at each step and the relatively large
Trotter error induced by the second-order decomposition of the time
evolution exponential. The stability of the full update strongly
depends on the choice of the corner transfer matrix renormalisation
growth and we are positive that improved methods -- potentially based
on better gauging\cite{evenbly18:_gauge} or a canonical-like form for
PEPS\cite{zaletel:_isomet} -- will be
introduced. Likewise, the large Trotter error encountered here could
be handled either by improved Trotter
decompositions\cite{barthel19:_optim_lie_trott_suzuk}, a hybrid
approach to evolve more than two sites
exactly\cite{hashizume18:_dynam_phase_trans_two_dimen} or by adapting
the variational ground-state search\cite{corboz16:_variat,
  vanderstraeten16:_gradien} to work as a two-dimensional
time-dependent variational principle.

\section*{Acknowledgements}

During the course of this project,
Ref.~\cite{czarnik19:_time}
similarly applied real-time evolution of iPEPS to the transverse field
Ising model.

\paragraph{Funding information}
The authors acknowledge funding through ERC Grant QUENOCOBA,
ERC-2016-ADG (Grant no. 742102) and by the Deutsche
Forschungsgemeinschaft (DFG, German Research Foundation) under
Germany’s Excellence Strategy -- EXC-2111 -- 390814868.

\end{document}